\documentclass[12pt]{article}

\usepackage[english]{babel}

\usepackage{epsf}
\usepackage{pazh}
\tightenlines

\sloppypar

\begin{document}

\title{\bf Mechanism for the Suppression of Intermediate-Mass Black Holes}

\author{\bf \hspace{-1.3cm} \ \
V.I. Dokuchaev\affilmark{1}, Yu.N. Eroshenko\affilmark{1},
S.G. Rubin\affilmark{2} and D.A. Samarchenko\affilmark{2}}

\affil{
{\it Institute for Nuclear Research, Russian Academy of Sciences, 
pr. 60-letiya Oktyabrya 7a, Moscow, 117319 Russia}$^1$\\
{\it Moscow Engineering Physics Institute - National Research Nuclear University,
Kashirskoe sh. 21, Moscow, 115409 Russia}$^2$}

\sloppypar \vspace{2mm} \noindent A model for the formation of supermassive primordial black holes in galactic nuclei with the
simultaneous suppression of the formation of intermediate-mass black holes is presented. A bimodal mass
function for black holes formed through phase transitions in a model with a ``Mexican hat'' potential has
been found. The classical motion of the phase of a complex scalar field during inflation has been taken into
account. Possible observational manifestations of primordial black holes in galaxies and constraints on
their number are discussed. 

\section*{INTRODUCTION}

There is almost no doubt that the nuclei of
most structured galaxies host supermassive black
holes (SMBHs) with masses $\sim10^6-10^{10}M_{\odot}$ (Antonucci
1993; Falcke and Hehl 2003; Ferrarese and
Ford 2004). There is also observational evidence for
the presence of intermediate-mass ($\sim10^3-10^5M_{\odot}$)
black holes (IMBHs) in galactic disks and globular
star clusters (Gebhardt et al. 2002; Lanzoni
et al. 2007; Noyola et al. 2008; Strohmayer and
Mushotzky 2009). In contrast to stellar-mass black
holes produced by the explosions of massive supernovae,
the origin of SMBHs and IMBHs remains
enigmatic. Several scenarios for the formation and
subsequent growth of SMBHs have been proposed
(for reviews and references, see Rees 1984, Begelman
et al. 1984, Dokuchaev 1991, and Dokuchaev
et al. 2007), which can arbitrarily divided into two
groups: astrophysical and cosmological. In astrophysical
scenarios, IMBHs and SMBHs result from
the dynamical evolution and gravitational collapses
of gas clouds or of central star clusters in galactic nuclei or from
multiple mergers of stellar black holes. Cosmological
scenarios suggest a very early formation of primordial
black holes (PBHs) at the radiation-dominated stage
from adiabatic density perturbations or through phase
transitions in the early Universe. A cosmological scenario
for the formation of massive PBHs through the
collapse of isothermal baryonic charge fluctuations
that arise during phase transitions at the inflationary
stage has also been proposed (Dolgov and Silk 1993;
Dolgov et al. 2009).

The currently most popular models are those of
hierarchical merging where SMBHs are formed by
multiple mergers of less massive IMBHs (Miller
and Colbert 2004; Kawakatu et al. 2005). The latter
could be formed, for example, during the collapses of
gas clouds in small dark matter halos at pregalactic
epochs, during hierarchical merging of stellar-mass
black holes that are the remnants of the first stars
(Islam et al. 2003), or during accretional swallowing
of the first massive stars by PBHs in their interiors
(Bambi et al. 2008).

One of the scenarios for the formation of PBHs
through phase transitions was developed by Rubin
et al. (2000, 2001a, 2001b), Khlopov and Rubin
(2004), Khlopov et al. (2005), and Dokuchaev
et al. (2005, 2008). Apart from the presence of
SMBHs in galactic nuclei, this scenario also predicts
the existence of IMBHs in galactic halos at a considerable
distance from their centers. The distinctive
features of the scenario are the formation of black
holes already at the radiation-dominated stage and
a characteristic spatial distribution of black holes in
clusters. Whereas ``astrophysical'' black holes are
formed in potential wells at the centers of dark matter
halos (Dokuchaev et al. 2007), PBHs are capable of
producing an induced dark matter halo around themselves
that appears as a dwarf galaxy (Dokuchaev and
Eroshenko 2001, 2003; Dokuchaev et al. 2008) with
a sharp increase in density toward the center. The
mergers of IMBHs formed through the mechanism
proposed by Rubin et al. (2000, 2001a, 2001b),
Khlopov and Rubin (2004), and Khlopov (2005) and
the formation of a peculiar class of galaxies and
quasars around primordial SMBHs were considered
by Dokuchaev et al. (2005, 2008).

It is hard to imagine that absolutely all IMBHs
will merge and produce central SMBHs in presentday
galaxies. It is more likely that only some fraction
of them, $\sim1$, will merge, while the remaining IMBHs
form a population of IMBHs in galactic halos. A
similar situation arises in many models with primordial
SMBHs. Indeed, if some mechanism predicts the
formation of one primordial SMBH in the bulk of one
galaxy, then it is hard to avoid the presence of a large
number of less massive IMBHs in the same galaxy.
In addition, even if SMBHs in galactic nuclei are not
related by their origin to IMBHs, the IMBH mass
functions tend to increase toward low masses.

Thus, a common feature of the above SMBH
formation mechanisms is the predicted abundance of
IMBHs. Nevertheless, observational data constrain
significantly the possible number of IMBHs in galaxies.
Only a small number of objects can be attributed
to IMBHswith some probability, for example, IMBHs
in some globular star clusters. However, as yet there
is no unequivocal proof of the presence of IMBHs
at their centers, because the available observational
evidence for their existence is questioned and comes
under justified criticism (Illingworth and King 1977;
McNamara et al. 2003; Baumgardt et al. 2003;
van der Marel and Anderson 2009). Besides, the
formation of IMBHs in globular clusters is severely
hampered by the heating of their central parts by
binary stars (hard pairs) formed during dissipative
two-body encounters of single stars (Ozernoy and
Dokuchaev 1982; Dokuchaev and Ozernoy 1982).

IMBHs could accrete gas and manifest themselves
as ultraluminous X-ray sources (ULXs), those
with a luminosity $L_X\ge10^{39}$~erg~s$^{-1}$. However, various
models have been proposed to explain ULXs
observed in the disks of some galaxies even without
invoking IMBHs (Roberts 2007; King 2008). We
have to say that the currently available observational
data give no unambiguous answer to the question
about the existence and abundance of IMBHs in
the Universe. There are only specific examples and
some evidence and theoretical arguments for their
presence. In this paper, we propose a model in which
the IMBH formation is suppressed under effective
SMBH formation in galactic nuclei. This scenario
can become topical if future observations will constrain
significantly the number of IMBHs. We propose
a modification of the mechanism initially suggested
by Rubin et al. (2001b) that allows not only
the presence of SMBHs at galactic centers but also
the suppression of IMBH formation to be explained.

\section*{THE FORMATION OF PRIMORDIAL BLACK HOLES WITH ALLOWANCE MADE
FOR THE CLASSICAL MOTION OF THE PHASE OF A COMPLEX SCALAR FIELD}

The subsequent discussion is based on the primordial
SMBH formation model developed by Rubin
et al. (2000, 2001a, 2001b), Khlopov and Rubin
(2004), Khlopov et al. (2005), and Dokuchaev
et al. (2005, 2008), who showed that if the inflaton
potential has a local maximum, then the SMBH formation
probability at the post-inflationary epoch is
high. These authors described in detail the formation
of a black hole through quantum field fluctuations
near this maximum. The consideration is based on
the shape of a potential known as a ``tilted Mexican
hat'', where the local maximum or, more precisely, the
saddle point is located at points $\theta =\pi , 2\pi ,
...$ ($\theta$ is the
phase of a complex scalar field). This potential, which
is used, for example, to describe inflation on a pseudo-
Nambu-Goldstone field (Dolgov and Freese 1995), is
\begin{equation}\label{potfi}
  V(|\Phi|)=\lambda [\Phi ^*\Phi-f^2/2]^{2}\quad ,
\end{equation}
with the minimum at $\langle \Phi \rangle =fe^{i\phi /f}/\sqrt{2}$. However, we
will emphasize that the field $\Phi$ under consideration is
additional to the inflaton one.

The problem is slightly simplified on energy scales
smaller than $f$, because the massive radial mode
of $\Phi$ ($m_{\rm rad}=\lambda ^{1/2}f$) can be discarded. The remaining
light degree of freedom is defined by the angular
variable $\theta=\phi /f$ that acts as a pseudo-Nambu-Goldstone boson (PNGB). Through spontaneous
$U(1)$ symmetry breaking on energy scales $\sim
\Lambda \ll f$, the PNGB potential acquires an additional term,
\begin{equation}\label{potlambda}
 V(\theta)=\Lambda ^4[1-\cos \theta] ,
\end{equation}
In the equation of motion for the field $\theta$ with potential
(1), (2),
\begin{equation}\label{teta1}
 \ddot{\theta}+3H\dot{\theta}+\frac{\Lambda ^4}{f^{2}}\sin\theta =0
\end{equation}
where the Hubble parameter $H$ remains constant
during inflation, because the field $\theta$ is not the inflaton
one. During inflation, the quantity $\theta$ changes slowly
and only after the end of the inflationary stage does $\theta$
execute rapid oscillations near the minima necessary
for efficient production of matter particles and heating
of the Universe. Depending on initial conditions, the
field $\theta$ can roll down to one of the minima, $\theta_{min}=0$ or
$\theta _{min}=2\pi$.

During the inflationary stage, when the field $\theta$ is
in the valley of the ``Mexican hat'' $\left| \Phi \right| =f/\sqrt{2}$, the
space is divided into many causally disconnected regions.
The values of $\theta$ in these regions slightly differ
due to quantum fluctuations. In some of them, the
field $\theta$ may turn out to be on the other side of the
maximum $\theta _{max}=\pi$ and after the end of inflation may
roll down to a value of the minimum different from
that to which most other regions tend. Thus, a Universe
that consists of chaotically distributed domains
with fields of 0 or $2\pi$ inside emerges after inflation.
The neighboring domains are separated by the field
walls whose subsequent evolution leads to the formation
of PBHs. The number of formed PBHs and
their mass depend strongly on the tilt of the potential
$\Lambda$ and the symmetry breaking scale $f$ at the onset
of inflation. Dokuchaev et al. (2005) chose model
parameters at which clusters consisting of initially
massive PBHs, the largest of which reached a mass
$\sim 4\cdot 10^7M_{\odot}$, were formed. Such massive PBHs can
subsequently serve as the nuclei of protogalaxies,
increasing their mass to $\sim 10^9M_{\odot}$ through accretion.
However, in addition to massive PBHs, a large number
of intermediate-mass ($\sim 10^4M_{\odot}-10^6M_{\odot}$) PBHs
are formed. As was discussed in the Introduction,
their number may turn out to be too large, which will
be in conflict with observational data. Themechanism
for the suppression of IMBH formation considered
here is based on the following almost obvious fact.
The farther the initial phase at which the present-day
Universe was formed from the maximum (in our case,
$\theta =\pi$), the smaller the volume of space will be filled
with a phase $\theta >\pi$. Indeed, after each e-fold, spatial
regions with phases advanced toward the maximum
of the potential by some small $\delta
\theta$ appear due to field
fluctuations. The time, the number of e folds needed
for the maximum of the potential to be reached, increases
sharply with increasing difference between
the initial phase and $\theta =\pi$. Therefore, the total number
of closed walls and, hence, black holes formed
from them also decreases sharply. In other words, the
dispersion of the phase distribution increases with
time and the farther the mean field from $\theta =\pi$, the
fewer the black holes will appear subsequently. If
the mean phase recedes from $\theta =\pi$ with time, then
the formation of IMBHs will be suppressed. This is
possible if the classical motion of the phase is taken
into account.

So far the classical motion of the phase $\theta$ of a
complex scalar field at the stage of inflation has been
neglected. This is true if the tilt of the potential $V(\theta)$
defined by the parameter $\Lambda$ is negligible. If, however,
this assumption does not hold, then, as a result of
the classical motion, the mean field recedes from the
maximum and the IMBH formation probability turns
out to be suppressed. Since themain idea of our paper
is take this effect into account, let us consider it in
more detail. We will measure the time in e-folds $N$
before the end of inflation. Let there be some region of
space appeared $N_r$ e-folds before the end of inflation
and filled with a field with phase $\theta_r (N_r)$.

We will be interested in the phase change with
time through both quantum fluctuations and classical
motion. An appropriate mathematical apparatus was
developed by Rey (1987), who proposed representing
the field as the sum of the classical term $\Theta$ and fluctuations
$\vartheta$ about it,
\begin{equation}\label{tt}
\theta = \Theta(t) +\vartheta , \quad \vartheta \ll \Theta
\end{equation}
The probability density $P_f (\vartheta)$ for detecting the fluctuational
part of the phase $\vartheta$ satisfies the Fokker-Planck equation, whose solution has the form of
a Gaussian distribution (Starobinsky 1982). Obviously,
the phase distribution of interest to us can be
obtained by the simple substitution $P(\theta)=P_f (\Theta -\theta)$.

The tilt of potential (1), (2) is small at the chosen
parameters $\Lambda =1.75H$ and $f=10H$ and we will take
it into account only when calculating the main, classical
contribution $\Theta(t)$ in (4). The small corrections
due to the fluctuations $\vartheta$ in (4) will be taken into
account by neglecting the tilt of the potential lest the
accuracy of the calculations be exceeded. In addition,
analytical results are known for the spatial phase
distribution (see, e.g., Starobinsky, 1982; Rey 1987).

In terms of e-folds, the probability of finding the
phase in the interval $(\theta ,
\delta \theta )$ at a certain point of space
is (Khlopov and Rubin 2004; Rubin et al. 2001b)
\begin{equation}
\delta P(\theta )=\frac{1}{\sqrt{2\pi \left( N_{r}-N\right) }}\exp \left[
-\frac{ \left( \theta _{r}(N)-\theta \right) ^{2}}{2\delta \theta
^{2}\left( N_{r}-N\right) }\right] ;\quad \delta \theta =\frac{H}{2\pi f}.
\end{equation}
Here, as was said above, we neglected the tilt of
the potential. The classical part of the phase $\Theta(t)$ is
represented as $\theta _{r}(N)$, where the number of e-folds is
$N=Ht$.

The condition for the appearance of one, and only
one, SMBH at time $N_0$
\begin{equation}
\delta P(\pi ) =\frac{1}{e^{3\left( N_{r}-N_{0}\right) }}
\end{equation}
indicates that black holes are formed when the extremum
of the potential $\theta =\pi$ is crossed. Now, from
the equation
\begin{equation}
e^{3\left( N_{r}-N_{0}\right) }=\sqrt{2\pi \left( N_{r}-N_0
\right) }\exp \left[ \frac{\left( \theta _{r}(N_0 )-\pi \right)
^{2}}{2\delta \theta ^{2}\left( N_{r}-N_0 \right) }\right],
\end{equation}
we will find the phase $\theta _{r}(N_{0})$ from which the spatial
region must be produced for the SMBH formation conditions to emerge $N_0$ e-folds before the end of
inflation,
\begin{equation}
\theta _{r}(N_{0})=\pi -\delta \theta \sqrt{\left(
N_{r}-N_{0}\right) \left[ 6\left( N_{r}-N_{0}\right) - \ln 2\pi
\left( N_{r}-N_{0}\right)  \right] }.
\end{equation}.

\begin{figure}[h]\label{theta}
\epsfxsize=17cm \hspace{-2cm}\epsffile{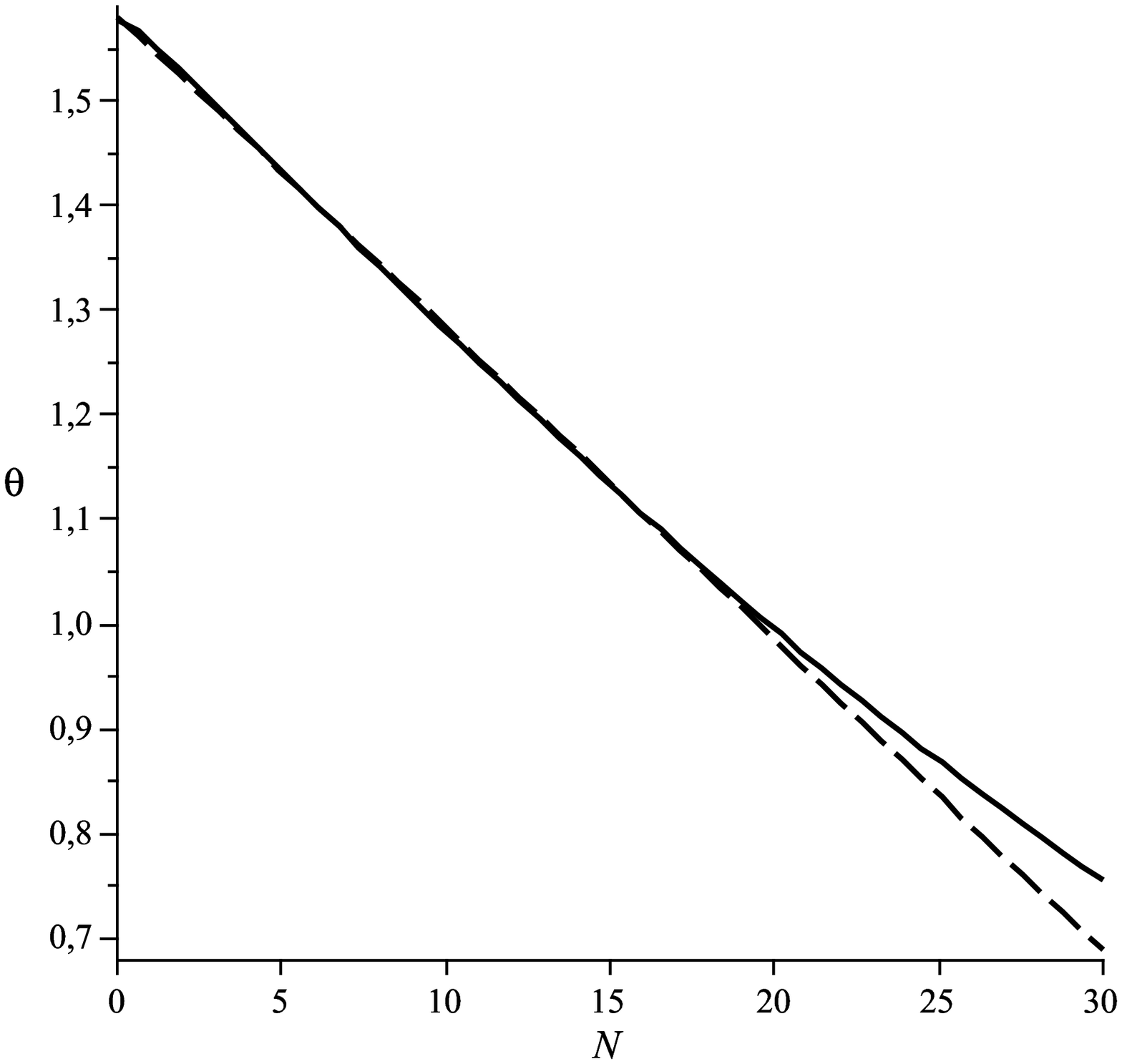}
 \caption{Dependence of $\theta$ on the number of e-folds at $\Lambda=1.75H$ and $f=10H$: the solid line and the dots represent the numerical calculation and the analytical calculation in the slow-rolling approximation, respectively; the fitting straight line
is indicated by dashes, its slope is $\alpha =0.028$.}
\end{figure}

Let us now determine the formation probability
of lower-mass black holes produced at an e fold $N$,
so that $N<N_0 <N_r$. Figure~1 show the dependence
of the phase $\theta$ on time expressed in e-folds
obtained by numerically solving Eq.~(3) and in the
slow-rolling approximation. The excellent agreement
indicates that the system is in the regime of slow
rolling. In the subsequent calculation of the probability
of quantum fluctuations in field $\theta$, it is appropriate
to use not the numerical solution of Eq.~(3) but its
linear fit. Since intermediate-mass PBHs are formed
from large domains emerging in a small number of efolds,
the solution of (3) is fitted in the initial segment
($N<20$). It is convenient to fit the dependence by a
linear function,
\begin{equation}
\theta _{r}(N)\simeq \theta _{r}(N_{0})-\alpha \cdot \left(
N-N_{0}\right)
\end{equation}
The probability for the emergence of IMBH formation
conditions during an e-fold number $N$ is then
\begin{equation}
\delta P(\pi )=\frac{1}{\sqrt{2\pi \left( N_{r}-N\right) }}\exp \left[ -\frac{%
\left( \theta _{r}(N_{0})-\alpha \cdot \left( N-N_{0}\right) -\pi
\right) ^{2}}{2\delta \theta ^{2}\left( N_{r}-N\right) }\right] .
\end{equation}
Given the number of causally disconnected regions
$exp[3(N-N_r)]$ formed since the formation of the
spatial region $N_r$, we can find the total number of
IMBHs in this region. A detailed description was
given by Khlopov and Rubin (2004), who, in contrast
to our case, disregarded the classical motion of the
field, i.e., $\alpha =0$.

\begin{figure}[h]\label{spectr}
\epsfxsize=17cm \hspace{-2cm}\epsffile{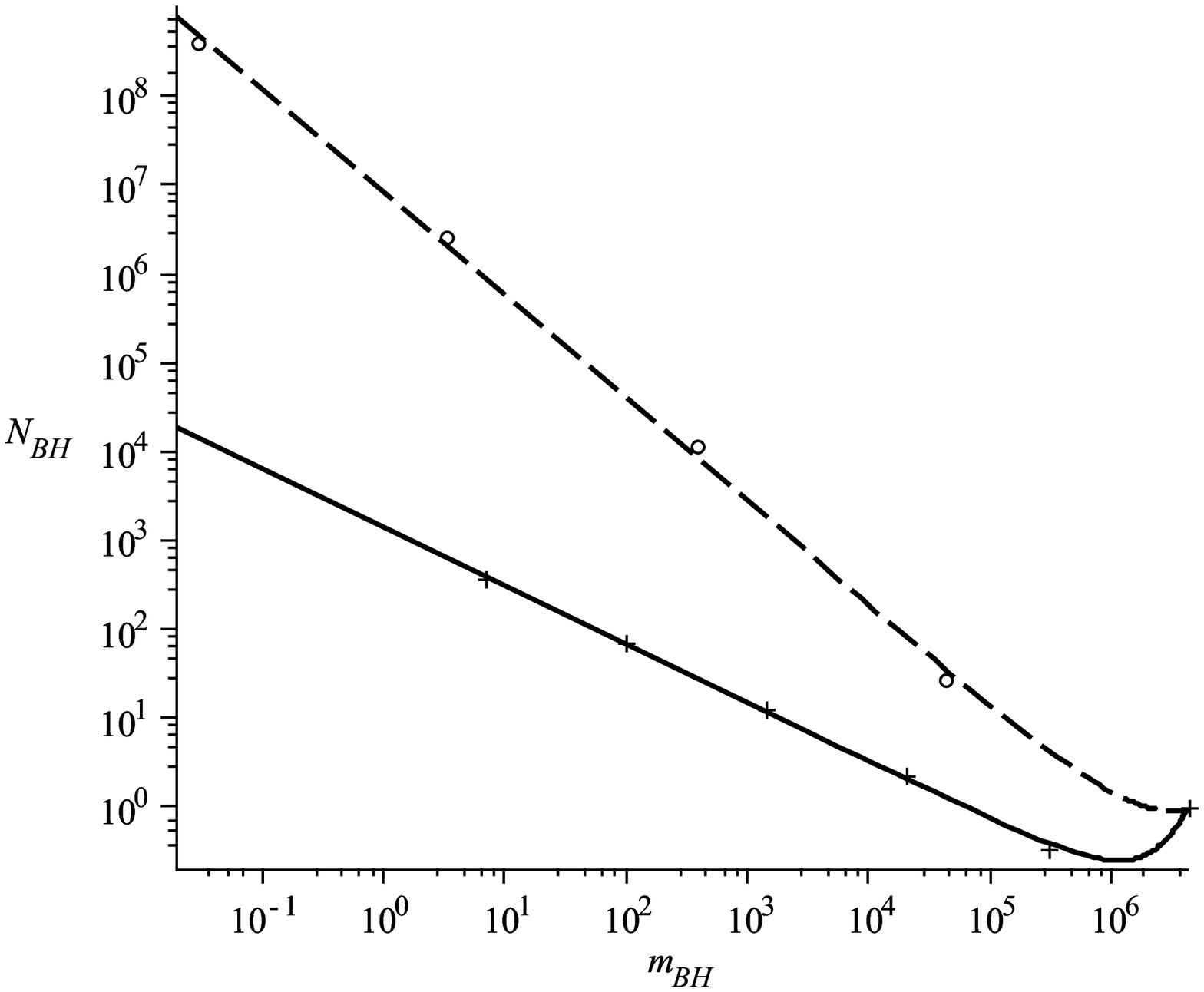}
\caption{PBH mass distribution: the solid line indicates the distribution for $\alpha =0.028$ and $\Lambda \approx1.75H$; the dashed line indicates
the distribution for $\alpha =0$.}
\end{figure}

Figure 2 presents the PBH mass distribution for
$\Lambda =1.75H$ and $f=10H$. As we see, the IMBH formation
actually turns out to be suppressed. The mass
function is bimodal in shape: the region of low-mass
black holes is separated by a dip from SMBHs.

\section*{ASTROPHYSICAL IMPLICATIONS}

In the preceding section, we showed that the PBH
mass spectrum could have two isolated regions. The
maximum at the highest masses gives SMBHs that
can become SMBHs in galactic nuclei, possibly, by
additionally increasing their mass through gas accretion.
Less massive IMBHs form a population of
black holes in galactic halos. Since their formation
is suppressed, these IMBHs are very rare and they
can be virtually unobservable. Let us discuss in more
detail the paths of black-hole evolution from these
two sets.

Under the influence of dynamical friction, a black
hole that was initially in the halo loses its angular
momentum and approaches the galactic center.
However, the dynamical friction mechanism is efficient
only for fairly massive objects, more specifically,
objects with a mass of more than $\le10^7M_{\odot}$ sink to
the galactic center in the Hubble time. An important
factor that contributes to the motion of black holes
to the galactic center is the formation of an induced
dark matter halo around the black hole (Dokuchaev
and Eroshenko 2001a, 2003; Mack et al. 2007). The
mass of the induced halo exceeds the black-hole mass
approximately by two orders of magnitude. Thus,
black holes with a mass $\ge10^5M_{\odot}$ surrounded by the
induced halo accumulate and merge at the galactic
center. If the internal structure of the induced halo
depends only on the PBH mass, then the boundary
of the induced halo is determined by the environment
in which the halo is formed. The neighboring density
fluctuations at a pregalactic stage will be an alternative
center of attraction for dark matter starting
from some distance from the black hole (Dokuchaev
and Eroshenko 2001a, 2003).

If a pair of IMBHs was formed at the galactic center
and if a third IMBH approached the pair under the
action of dynamical friction, then the slingshot effect
is possible: the pair becomes even closer as energy is
exchanged during the gravitational interaction, while
the third IMBHacquires a high velocity and is ejected
away from the galactic center (Haehnelt and Kauffmann
2002). The detection of such ejected IMBHs
would serve as evidence for the formation of SMBHs
from merging lower-mass IMBHs. Another method
for testing various SMBH formation models is to
search for gravitational wave bursts from black-hole
mergers. For example, in the model from Dokuchaev
et al. (2008), IMBHs in central clusters in galactic
nuclei collide with one another with the generation
of gravitational wave bursts (Dokuchaev et al. 2009),
with the burst redshift distribution differing from that
provided by ``astrophysical'' scenarios for the origin
of SMBHs. Therefore, observations with the LISA
telescope planned to be launched can provide a choice
between various scenarios for the origin of SMBHs
(Dokuchaev et al. 2009).

In an alternative scenario, for example, that considered
here, the central black hole initially had a high
mass. In the hierarchical picture of galaxy formation,
a SMBH fell into a large galaxy together with one of
the small protogalaxies and rapidly sank to its dynamical
center through dynamical friction. The advantage
of the model with massive primordial SMBHs is that
it can successfully explain the presence of quasars at
high redshifts (Dokuchaev et al. 2005).

Lower-mass ($\le10^5M_{\odot}$) black holes together with
the surrounding induced dark matter halos have no
time to sink to the center and remain in the galactic
halo. They can be observable as ultraluminous Xray
sources when passing through dense molecular
clouds in the disk. Since the gas density in the galactic
halo is low, accretion directly from the halo will
not give a significant X-ray flux. However, the case
where an IMBH moving in the halo carries a small
induced dark matter halo and, most importantly, a
cloud of baryons and an accretion disk is possible.
In this case, the X-ray emission from the IMBH
will be determined by the accretion of baryons from
the induced halo. This scenario appears problematic,
because no cooling of baryons, in particular, no star
formation, takes place in low-mass halos. For this
reason, there is no flow of gas onto the central IMBH.
This also implies that the observed ultraluminous Xray
sources, candidates for the role of IMBHs, are
present only in galactic disks.

Let us discuss the growth of IMBH masses
through accretion in the time of their passages
through the galactic disks. For Bondi-Hoyle accretion,
a black hole has a luminosity $L_X=4\pi\eta c^2G^2M^2\rho v^{-3}$ (Mii and Totani 2005; Mapelli
et al. 2006), where $v$ is the velocity with which
the black hole moves through a gas with density $\rho$
and $\eta$ is the mass-to-energy conversion efficiency.
This parameter is $\eta\simeq0.06$ for disk accretion onto a
Schwarzschild black hole and reaches $\eta\simeq 0.42$
accretion onto an extreme Kerr black hole. Typically,
$\eta\simeq 0.1$. In this case, the black hole mass increases
with the characteristic time $t_{\rm BH}=(4\pi G^2M\rho v^{-3})^{-1}$.
If the luminosity is assumed to correspond to the
luminosity of observed ultraluminous X-ray sources,
then
\begin{equation}
t_{\rm
BH}\simeq6\times10^{10}\left(\frac{M}{10^4M_{\odot}}
\right)\left(\frac{L_X}{10^{39}\mbox{~erg~s$^{-1}$}}\right)^{-1}\mbox{~yr}.
\end{equation}
Thus, an IMBH has no time to increase appreciably
itsmass in the lifetime of theUniverse even if it always
moves in the disk. A larger mass growth could be
expected for the Eddington luminosity of a black hole
at rest in a molecular cloud. However, the probability
of an IMBH being captured into the galactic disk is
very low. The probability of an IMBH being directly in
a cloud of cold gas is even lower, because molecular
clouds occupy only a disk volume fraction $f_{\rm MC}\simeq0.017$ (Mapelli et al. 2006). Consequently, no blackhole mass redistribution occurs and the region of the
IMBH mass function remains suppressed.

\section*{CONCLUSIONS}

Here, we presented a new SMBH formation
mechanism in which the formation of IMBHs is
suppressed. Whereas the existence of SMBHs at
galactic centers is already beyond doubt, the currently
available observational data provide only circumstantial
evidence for the existence of IMBHs. Since
in some models for the formation of black holes
their mass spectrum is close to a power law, the
formation of SMBHs is inevitably accompanied by
the formation of a large number of IMBHs. In the
model we developed, this does not happen. On the
contrary, the formation of IMBHs is suppressed
through the classical motion of the phase of a scalar
field additional to the inflaton. If detailed astronomical
observations will establish that the number of IMBHs
in the Universe is very small, then this fact can be
explained in the presented model.

\section*{ACKNOWLEDGMENTS}
This study was supported in part by the Russian
Foundation for Basic Research (project nos. 09-02-00677 and 10-02-00635), the Program of the
President of Russia for Support of Leading Scientific
Schools (project 3517.2010.2), and the Russian Agency for Science
(State contract 02.740.11.5092).

\section*{REFERENCES}

1. R. R. J. Antonucci, Ann. Rev. Astron. Astrophys. 31,
473 (1993).

2. C. Bambi, D. Spolyar, A. D. Dolgov, et al.,
arXiv:astro-ph0812.0585v2 (2008).

3. H. Baumgardt, J. Makino, P. Hut, et al., Astrophys. J.
589, L25 (2003).

4. M. C. Begelman, R. D. Blandford, and M. J. Rees,
Rev. Mod. Phys. 56, 255 (1984).

5. V. Dokuchaev, Yu. Eroshenko, and S. Rubin,
arXiv:astro-ph0709.0070v2 (2007).

6. V. Dokuchaev, Yu. Eroshenko, and S. Rubin, Grav.
Cosmol. 11, 99 (2005).

7. V. I. Dokuchaev, Yu. N. Eroshenko, and S. G. Rubin,
Astron. Zh. 85, 867 (2008) [Astron. Rep. 52, 779
(2008)].

8. V. I. Dokuchaev and Yu. N. Eroshenko, Pis’ma Astron.
Zh. 27, 883 (2001) [Astron. Lett. 27, 759
(2001)].

9. V. I. Dokuchaev and Yu. N. Eroshenko, Astron. Astrophys.
Trans. 22, 727 (2003).

10. V. I. Dokuchaev and L. M. Ozernoy, Astron. Astrophys.
111, 16 (1982).

11. V. I. Dokuchaev, Yu. N. Eroshenko, and S. G. Rubin,
Pis’ma Astron. Zh. 35, 1 (2009) [Astron. Lett. 35, 143
(2009)].

12. V. I. Dokuchaev, Usp. Fiz. Nauk 161, 1 (1991) [Sov.
Phys. Usp. 34, 447 (1991)].

13. A. Dolgov and K. Freese, Phys. Rev. D 51, 2693
(1995).

14. A. D. Dolgov, M. Kawasaki, and N. Kevlishvili, Nucl.
Phys. B 807, 229 (2009).

15. A.Dolgov and L. Silk,Phys. Rev. D 47, 4244 (1993).
16. L. Ferrarese and H. Ford, arXiv:astro-ph/0411247
(2004).

17. The Galactic Black Holes, Series in High Energy
Physics, Cosmology, and Gravitation, Ed. by
H. Falcke and F.W. Hehl (IOP Publ., London, 2003).

18. K. Gebhardt, R.M. Rich, and L. C. Ho, Astrophys. J.
578, L41 (2002).

19. M. G. Haehnelt and G. Kauffmann,Mon. Not. R. Astron.
Soc. 336, L61 (2002).

20. G. Illingworth and I. R. King, Astrophys. J. 218, L109
(1977).

21. R. R. Islam, J. E. Taylor, and J. Silk, Mon. Not.
R. Astron. Soc. 340, 647 (2003).

22. N. Kawakatu, T. R. Saitoh, and K. Wada, Astrophys.
J. 628, 129 (2005).

23. M. Yu. Khlopov and S. G. Rubin, Cosmological Pattern
of Microphysics in the Inflationary Universe
(Kluwer Acad., Dordrecht, 2004), v. 144.

24. M. Yu. Khlopov, S. G. Rubin, and A. S. Sakharov,
Astropart. Phys. 23, 265 (2005).

25. A. King,Mon.Not. R. Astron. Soc. 385, L113 (2008).

26. B. Lanzoni, E. Dalessandro, F. R. Ferraro, et al., Astrophys.
J. 668, L139 (2007).

27. K. J.Mack, J. P.Ostriker, and M. Ricotti, Astrophys.
J. 665, 1277 (2007).

28. M. Mapelli, A. Ferrara, and N. Rea, Mon. Not. R. Astron.
Soc. 368, 1340 (2006).

29. R. P. van derMarel and J. Anderson, arXiv:0905.0638
(2009).

30. B. J. McNamara, T. E. Harrison, and J. Anderson,
Astrophys. J. 595, 187 (2003).

31. H.Mii and T. Totani, Astrophys. J. 628, 873 (2005).

32. M. C. Miller and E. J. M. Colbert, Int. J. Mod. Phys.
D 13, 1 (2004).

33. E. Noyola, K. Gebhardt, and M. Bergmann, Astrophys.
J. 676, 1008 (2008).

34. L. M. Ozernoy and V. I. Dokuchaev, Astron. Astrophys.
111, 1 (1982).

35. M. J. Rees, Ann. Rev. Astron. Astrophys. 22, 471
(1984).

36. S.-J. Rey, Nucl. Phys. B 284, 706 (1987).

37. T. P. Roberts, Astrophys. Space Sci. 311, 203 (2007).

38. S. G. Rubin, M. Y. Khlopov, and A. S. Sakharov,Grav.
Cosmol. S 6, 51 (2000).

39. S. G. Rubin, A. S. Sakharov, and M. Yu. Khlopov,
Zh. E´ ksp. Teor. Fiz. 119, 1067 (2001) [JETP 92, 921
(2001)].

40. S. G. Rubin, A. S. Sakharov, and M. Yu. Khlopov,
Zh. E´ ksp. Teor. Fiz. 119, 1067 (2001) [JETP 92, 921
(2001)].

41. A. Starobinsky, Phys. Lett. B 117, 175 (1982).

42. T. E. Strohmayer and R. F. Mushotzky, Astrophys. J.
(in press);

43. T. E. Strohmayer, arXiv:astro-ph.HE0911.1339
(2009).

\end{document}